\documentclass[pra,twocolumn]{revtex4}
\usepackage{epsfig}
\usepackage{latexsym,exscale}
\usepackage{bm}
\usepackage{amsmath,amssymb,amsfonts}
\def\be{\begin{eqnarray}} 
\def\ee{\end{eqnarray}}

\def\bp{\bm{p}}

\def\bx{\bm{x}}

\def\vecu{\bm{u}}
\newcommand{\Fd}{F^{\dagger}}

\newcommand{\1}{{\uparrow}}
\newcommand{\2}{{\downarrow}}

\begin{document}
\title{Anomalous specific heat jump in a two-component ultracold Fermi gas
}
\author{Armen Sedrakian$^1$,  Herbert M\"uther$^1$, and Artur Polls$^2$}
\affiliation{$^1$Institute for Theoretical Physics, 
T\"ubingen University, D-72076 T\"ubingen, Germany\\
$^2$Departament d'Estructura i Constituents de la Mat\`eria,
Universitat de Barcelona, E-08028 Barcelona, Spain\\
}


\begin{abstract}
The thermodynamic functions of a Fermi gas with spin population 
imbalance are studied in the temperature-asymmetry plane in the 
BCS limit.
The low temperature domain is characterized by anomalous 
enhancement of the entropy and the specific heat above their 
values in the unpaired state, decrease of the gap and eventual  
unpairing phase transition as the temperature is lowered.
The unpairing phase transition induces a second jump in the 
specific heat, which can be measured in 
calorimetric experiments. While the superfluid is 
unstable against a supercurrent carrying state, it may 
sustain a metastable state if cooled adiabatically down from the 
stable high-temperature domain. In the latter domain 
the temperature dependence of the gap and  related functions is 
analogous to the predictions of the BCS theory.
\end{abstract}

\maketitle

Recent experiments~\cite{MIT,RICE} on ultracold dilute gases of fermionic 
atoms trapped unequal number of fermions in two different hyperfine states. 
These experiments started addressing some of the long standing problems 
in the theory of asymmetric superconductors (ASC) that are of interest in 
variety of fields including metallic superconductors~\cite{LO,FF}, nuclear 
systems~\cite{SEDRAKIAN,NOZIERES,MUTHER} and high 
density QCD~\cite{ALFORD,ALFORD2,BOWERS,NARDULLI}. The 
unprecedented control over the many-body systems achieved 
in the experiments with ultracold dilute fermions combined 
with the possibility of tuning the interactions via the 
Feshbach resonance mechanism provide for the first time a realistic
perspective of testing the predictions of the theories of ASC in the
context of dilute fermionic systems. The realizations of various 
phases of ASC of dilute fermions have been intensively studied on 
the theoretical front; the simplest realizations are the 
isotropic, homogeneous phases that are characterized either by 
a Zeeman splitting of Fermi-levels~\cite{G1,G2,G3} or
by pairing between light and heavy fermions~\cite{G4a,G4b}. 
At large asymmetries the  phases with broken space  
symmetries~\cite{G5a,G7,G5b,G8,G9,G11,G16,L_HE} and the
mixed phases~\cite{G6,G10}  become energetically 
more favorable. Alternatives include pairing in higher angular 
momentum states~\cite{G12a,G12b}. Finite size and trap geometry 
introduce an additional complication to the problem and may qualitatively 
affect the comparison between the theory and 
experiment~\cite{CHEVY,DESILVA}. 
A number of related problems of interest are
the nature of phase transitions between the various phases
and their relation to the topology of 
Fermi-surfaces~\cite{G13a,G13b} as well as the features of 
the BCS-BEC crossover~\cite{G14,G17,PAO} under population 
imbalance.

The population asymmetry in ASC can be characterized either in 
terms of the difference (mismatch) in the chemical potentials or 
the difference in the densities of the species. The first case 
arises when the ``chemical'' equilibrium between populations admits
transmutation between the different spin states, as e.~g. under
the equilibrium with respect to the weak interactions in cold
dense hadronic/quark matter. We shall specify our discussion 
from the outset to the second case and assume that the total number of 
fermions is fixed and the asymmetry is maintained with respect to the 
{\it number densities} of different species, as is the case in the 
experiments on ultracold fermions. 

The aim of this {\it Letter} is the study of the temperature-asymmetry
phase diagram of an ultracold Fermi gas with pairing between two 
unequally populated hyperfine states in the BCS regime.  
We propose that calorimetric experiments, which are within the 
current experimental capabilities~\cite{JET}, can 
test the specific features of the phase diagram discussed 
below. The critical temperature of metallic superconductors,
according to the Bardeen-Cooper-Schrieffer 
(BCS) theory, is given by $T_c = 1.14~\omega_D~ e^{-1/V}$, 
where $\omega_D$ is the Debye frequency and 
$V$ is the dimensionless interaction. The parameters in this equation 
can be determined from a single calorimetric experiment; $T_c$ is 
determined from the position of the jump in the specific heat, 
and $\omega_D$ from the slope of the specific heat over $T^3$ 
in the limit $T\to 0$. The case of ASC is complicated by the 
fact that the  asymmetry leads to a loss of coherence at low temperatures 
and the critical temperature and pairing gap become complicated functions 
of spin imbalance. This paper studies 
the impact of the asymmetry induced decoherence  
on the thermodynamics of ultracold gases; in particular we show 
that two jumps in the specific heat of ASC are possible if the 
asymmetry is large, albeit the second anomalous jump occurs withing 
the temperature domain where the superfluid is in a metastable state. 
The anomalous jump in the specific heat is a manifestation
of the {\it reentrance effect}, i.~e., the restoration of 
pair-correlations at finite temperatures 
$T\le T_c$~\cite{SEDRAKIAN,SLA,BALIAN}.
Below, we shall confine ourselves to the case of infinitely 
extended systems.  Since the experiments are carried out in
finite geometries, finite-size corrections need to be taken into 
account in a more complete analysis. Recent 
experiments~\cite{MIT,RICE} which have measured the density 
profiles of trapped gases with population imbalance demonstrate 
the importance of these effects in determining the ground state 
structure of the condensate~\cite{CHEVY,DESILVA}. Related 
work on finite temperature phase diagram of ASC appeared 
in refs.~\cite{LEVIN,HE} while our work was in preparation/revision.

We approximate the pairing interaction  by a zero-range force,
which is characterized by the $s$-wave scattering length $a_S$.
Our discussion is specified to the case where two hyperfine states 
of $^6$Li are populated (the scattering length
in units of the Bohr radius is  $a_S/a_B = -2160$, but can be 
varied at will via Feshbach resonance mechanism). The fermion
masses are assumed to be equal; the extension to the case of unequal masses 
(as would be the case in the mixtures of $^6$Li and $^{40}$K) 
is straightforward.

We consider a uniform gas of fermionic atoms in two 
hyperfine states (spins) labeled as $\1$ and $\2$; the interaction 
Hamiltonian is 
\begin{eqnarray}\label{HAMILTON}
\hat H_{\rm int} = 
- V\sum_{\alpha\beta}\int d^3 x \hat \psi^{\dagger}_{\alpha}
(\bx)  \hat \psi^{\dagger}_{\beta}(\bx) 
\hat \psi_{\beta}(\bx) \hat\psi_{\alpha}(\bx),\nonumber
\end{eqnarray}
where $\hat \psi^{\dagger}_{\alpha}(\bx)$ and $\hat \psi_{\alpha}(\bx)$ are the
creation and annihilation operators of a state at the space point specified 
by the position vector $\bx$ and spin $\alpha (= \uparrow,\downarrow)$
and  $V$ is the two-body bare contact interaction.  
The normal and anomalous propagators are~\cite{G5b}
$G_{\1\2}(p) = \omega +{\cal E}_S\pm 
{\cal E}_A/ D$, $\Fd(p) = \Delta^{\dagger}/ D$ , where
$D \equiv (\omega-{\cal E}_A)^2-{\cal E}_S^2-\Delta^2$ and 
${\cal E}_S = (\varepsilon_{p\1}+\varepsilon_{p\2})/2$ and 
${\cal E}_A = (\varepsilon_{p\1}-\varepsilon_{p\2})/2$ are, 
respectively, the parts of the quasiparticle spectrum which are 
symmetric and antisymmetric under time-reversal operation, 
$\varepsilon_{p\1}, \varepsilon_{p\2},$ are the single particle energies
in states $\1$ and $\2$. The dispersion relation of the quasiparticles 
in the paired state is
$\omega_{\1/\2}  = {\cal E}_A \pm \sqrt{{\cal E}_S^2+\Delta^2}.$
The limit ${\cal E}_A \to 0$ corresponds to the case of equal number 
of spin-up and down particles. The explicit form of the symmetric 
and antisymmetric (under time-reversal) parts of the quasiparticle 
spectrum are ${\cal E}_S = (p^2/m - \mu_{\1}^*-\mu_{\2}^*)/2$ and 
${\cal E}_A =( - \mu_{\1}^*+\mu_{\2}^*)/2$, 
where $m$ is the atom's bare mass, $\bp$ is the relative momenta 
of fermions bound in a Cooper pair in the center-of-mass frame at rest, 
the effective chemical potentials $\mu_{\1/\2}^* = \mu_{\1/\2} -
\Sigma_{\2/\1}$ include the constant shift due to the self-energy 
$\Sigma_{\1/\2} =  T_0 \rho_{\2/\1}$, where $T_0$ is the two-body scattering 
$T$-matrix. The gap equation is 
\be\label{GAP2} 
\frac{2}{U_0} = \int_0^{\Lambda} \frac{1}
{\sqrt{{\cal E}_S^2(p)+\Delta^2}}[ f(\omega_\1)-f(\omega_\2) ]
\frac{p^2dp}{(2\pi)^2},
\ee
where $U_0$ is the strength of a contact interaction and $\Lambda$ is an 
ultraviolet cut-off~(for details see~\cite{G5b}). 
The occupation probabilities of species are given by  
\begin{eqnarray}\label{OCCUP} 
n_{\1/\2}(\bp) = 
 u_{\bp}^2\left[f(\omega_{\1/\2})-f(-\omega_{\2/\1})\right]
+ f(-\omega_{\2/\1}), 
\end{eqnarray}
where $u_{\bp}^2 = 1/2+{\cal E}_S /(2\sqrt{{\cal E}_S^2+\Delta^2})$;
these are normalized to the densities of species
$\rho_{\1/\2}=\sum_{\bp} n_{\1/\2}(\bp)$. 
The free-energy is ${\cal F} = \sum_{\bp} (\epsilon_{\bp\1}
n_{\bp,\1}+\epsilon_{\bp\2}n_{\bp,\2}) -U_0^{-1}\Delta^2 - TS_S$. 
The free-energy of the  normal state 
follows by setting in this expressions $\Delta = 0 =\delta\epsilon$.
The entropy of ASC is defined in terms 
of the temperature derivative of the free-energy 
$S_S = -\partial {\cal F}/\partial T$.
The specific heat follows as $C_V = T (\partial S_S/\partial T) = 
-T (\partial^2 {\cal F}/\partial T^2)$. 
\begin{figure}
\psfig{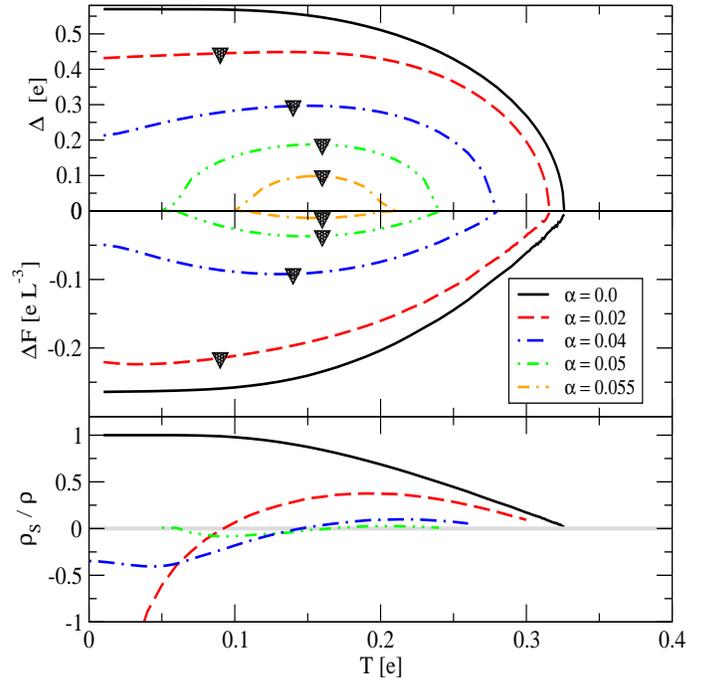}
\vskip 1. cm
\caption{Dependence of the pairing gap {\it (upper panel)} 
the free-energy difference {\it (middle panel)} 
and the superfluid density {\it (lower panel)} on the 
temperature for several values of the density asymmetry.
The instability domain ($\rho_s < 0$) lies to the left 
of triangles.}
\label{fig:GAP}
\end{figure}

The local stability requires that the free energy is a convex 
function of the appropriate variables and it has been established 
that homogeneous ASC could become unstable in this 
sense~\cite{G4b,G6,L_HE,G14,PAO,G13b,LEVIN,K_FUKUSHIMA,SHOVKOVY}.
Specifically, 
(A) the system is unstable against phase separation unless the 
curvature matrix $\chi_{ij} = \partial ^2{\cal F_S}/\partial\mu_i
\partial\mu_j$ is positive definite. This implies that
either  the eigenvalues 
$\lambda_{1,2} = {\rm Tr}\chi_{ij}/2 \pm 
\sqrt{({\rm Tr}\chi_{ij})^2-4{\rm Det}\chi_{ij}}/2 \ge 0$
or, equivalently, $\chi_{\1\1}>0$, Det $\chi_{ij}>0$ (Sylvester criterion). 
Further, (B) the condition $\partial ^2{\cal F_S}/
\partial\Delta^2>0$ needs to be fulfilled. 
Finally, (C) the system may become unstable against
spontaneous generation of currents (formation of the LOFF
phase) when  $\chi_{q} = \partial ^2{\cal F_S}/\partial q^2  <0$, 
where $q$ is the center-of-mass of momentum of a Cooper pair. 
The latter instability manifests itself in the 
negative superfluid density, $\rho_s = m^2\chi(q)\vert_{q=0}$,
and purely imaginary Meissner mass.
For small quasiparticle velocity $\vecu$ the leading order 
contribution to the ratio of the superfluid  to the total density is
\cite{WALECKA}
\be 
\frac{\rho_s}{\rho}(T) = 1 -\frac{1}{\rho m}\int\frac{dp p^4}{6\pi^2}
\left[\frac{\partial f(\omega_{\2})}{\partial\omega_{\2}}\Bigg|_{\vecu=0}
+\frac{\partial f(\omega_{\2})}{\partial\omega_{\1}}
\Bigg|_{\vecu=0}\right].
\ee
\begin{figure}[t]
\epsfig{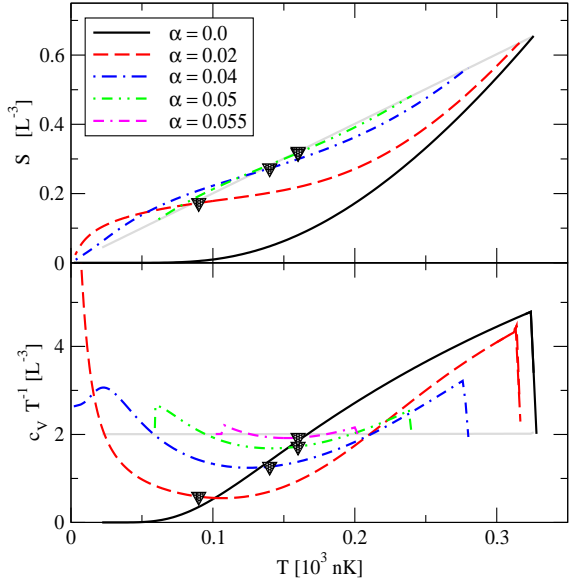}
\vskip 1. cm
\caption{Entropy {\it (upper panel)} and the heat capacity
{\it (lower panel)} as a function of the temperature for several values of 
the density asymmetry. Note that for $\alpha = 0.02$, 
$c_V(T=0) =  4.5$ is finite.
The grey lines show the values in the normal  state.}
\label{fig:ENTROPY}
\end{figure}
We fix the density of $^6$Li atoms at  
$\rho=3.8\times10^{12}$ cm$^{-3}$, which in the case of a
spatially uniform and spin-symmetric system at $T=0$
corresponds to the Fermi momentum $k_F = 4.83\times10^4$ cm$^{-1}$
and $k_Fa_S = -0.558$. Since we work in the
BCS regime the conditions  (A) and (B) are fulfilled 
(in agreement with ref.~\cite{LEVIN}).
Figure 1 displays the  $\Delta(T)$, $\Delta{\cal F}(T) = 
{\cal F}_N - {\cal F}_S$ and $\rho_s(T)/\rho$ functions 
for constant asymmetry 
$\alpha = (\rho_{\1}-\rho_{\2})/(\rho_{\1}+\rho_{\2})$. The length 
and energy are measured below in units of $L = 10^{-4}$ cm and 
$e = 10^2$ nK, unless explicitly specified.
Near the critical temperature the  asymptotic behavior  of the 
pairing gap for $T\to T_c$ is described, 
to leading order in $(\Delta/T)^2$,
by the BCS-type relation $\Delta(\alpha) \sim
[T_c(\alpha)(T_c(\alpha)-T)]^{1/2}$, i.~e., the high 
temperature portions of the $\Delta(T)$ curves are self-similar.
The low-temperature BCS asymptotics $\Delta(T)\sim \Delta(0) - 
\sqrt{2\pi T \Delta(0)}e^{-\Delta(0)/T}$ is qualitatively modified even 
at small asymmetries, since the gap decreases as $T\to 0$, 
instead of staying constant.  At large asymmetries 
(e.g. $\alpha \ge 0.04$) the reentrance effect sets in: 
the gap is non-zero only in a finite 
domain of temperatures bounded by two critical temperatures. The 
physical origin of the upper critical temperature is analogous to the
BCS case, where the pairing correlations are destroyed by thermal 
fluctuations. The lower critical temperature is due to the loss 
of coherence induced by the asymmetric population and is specific 
to ASC. At large asymmetries the temperature dependence of the gap 
is $\Delta(\alpha) \sim [T^*_c(\alpha)(T-T^*_c(\alpha))]^{1/2}$, 
where $T_c^*$ is the lower critical temperature. The 
free-energy plots (Fig.~\ref{fig:GAP})  reflect the 
temperature dependence of the condensation energy, which scales 
as $\Delta^2 (\alpha, T)$; the temperature dependence and asymptotics
of $\Delta{\cal F}(T)$ is understandable in terms of this scaling.
In the temperature domain where $\rho_s(T) < 0$, the homogeneous ASC
is metastable; there exists a lower extremum (perhaps minimum) 
of the free energy
corresponding to the current carrying LOFF phase. Nevertheless 
if prepared at high enough (but  $T< T_c$) temperature and cooled down
adiabatically, the metastable phase can be sustained long enough 
to carry out measurements.
\begin{figure}[t]
\epsfig{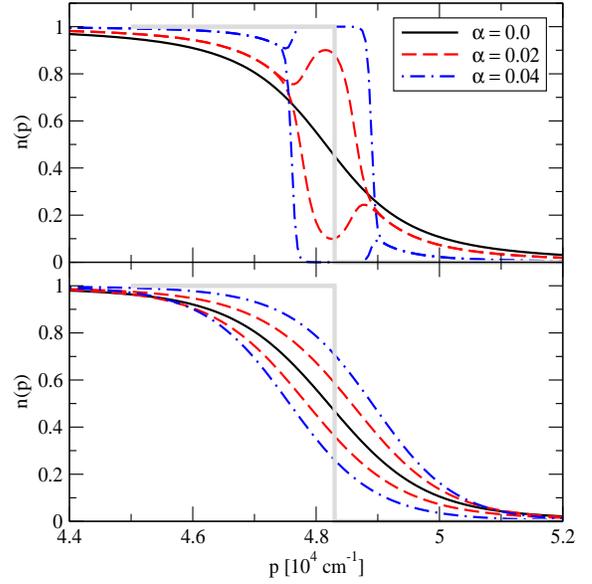}
\vskip 1. cm
\caption{Occupation probabilities of the majority and minority 
components as a function of the momentum for several asymmetries 
at temperatures $T = 1$ nK ({\it upper panel}) and $T = 25$ nK
({\it lower panel}). All lines correspond to stable states, except
$\alpha \neq 0$ lines in the upper panel. 
The grey line shows the same for the unpaired symmetric state at $T=0$.}
\label{fig:OCCUP}
\end{figure}
The temperature dependence of the entropy and specific heat 
(more precisely $C_V/T$) is shown in Fig.~\ref{fig:ENTROPY}. 
At $T\to T_c^{-}$ the entropy  scales linearly with temperature,  
$S_S \propto T-T_c$, with asymmetry dependent slope. 
In the low-temperature metastable region the superfluid entropy 
is anomalous, since its absolute magnitude is larger than 
the entropy of the normal state, i.~e., the superfluid appears 
to be less ordered than the unpaired state. The temperature for 
the onset of anomalous regime ($S_S > S_N$) coincides with that 
for the onset of instability within our numerical accuracy.
              
The ratio  $\Delta C/T_c$, where $\Delta C = C_S-C_N$ is the  jump 
in the specific heat at the critical temperature, depends only on the 
density of states and is a universal characteristic of a system.
The jump itself is a characteristic feature of a second order phase
transition that allows to locate $T_c$ experimentally. 
There is a second (anomalous) jump in the specific heat associated with  
the reentrance effect at the lower critical  temperature $T_c^*$, 
which lies within the metastable domain.
At $T\to T_c^{-}$ the  specific 
heat scales as $C_V \propto \vert T- T_c\vert$. 
Its low-temperature asymptotics  differs dramatically from the predictions
of BCS theory, where the superfluid ``thermal inertia'' is small 
compared to the normal case. The specific heat of ASC is larger 
than that of the normal state at sufficiently low temperature for 
{\it any} asymmetry. Calorimetric experiments 
aimed at measuring the specific heat of ASC can locate the critical 
temperature and observe the reentrance effect  through 
the second (anomalous)  jump  $\Delta C_V$, if the system can 
be maintained long enough in the metastable state.
       
The occupation probability of the majority and minority components
are shown in Fig.~\ref{fig:OCCUP} for the high- and low-temperature 
regions of pairing. The remarkable difference between these two 
cases arises from the fact that in the low-temperature regime 
the minority component is excluded from the region around the 
Fermi-momentum (``blocking region''~\cite{NOZIERES} or 
``breach''~\cite{G4a,G4b}). The depletion is large for large asymmetries.
In contrast, the high-temperature regime does not feature 
a depletion region and the occupation are smooth functions of the
momenta for arbitrary asymmetries.

The finite range of interactions and finite size of the 
systems are not likely to modify the conclusions reached here. 
Indeed, reentrance behavior(s) have been predicted in paired nuclear
systems which are characterized by complex finite range 
interactions~\cite{SLA,SEDRAKIAN} and ultra-small metallic grains, 
which contain a small number of fermions and a single (odd) unpaired 
particle~\cite{BALIAN}.


We are grateful to Jorge Dukelsky, Jordi Mur-Petit,  Peter Schuck and 
Claus Zimmerman for useful interactions. This work was in part 
supported by the SFB 382 of the DFG (Germany) and 
Grants No. FIS2005-03142 (MEC, Spain and FEDER) and No.
2005SGR-00343 (Generalitat de Catalunya).

\end{document}